\documentstyle[12pt]{article}
\makeatletter
\renewcommand\thesection{\@Roman\c@section}
\renewcommand\thesubsection{\thesection.\@arabic\c@subsection}
\makeatother

\begin{document}
\begin{titlepage}
\begin{flushright}
math-ph/9905002
\end{flushright}
\vskip.3in

\begin{center}
{\Large \bf Quasi-Spin Graded-Fermion Formalism and
       $gl(m|n)\downarrow osp(m|n)$ Branching Rules}
\vskip.3in
{\large Mark D. Gould} and {\large Yao-Zhong Zhang}
\vskip.2in
{\em Department of Mathematics, University of Queensland, Brisbane,
     Qld 4072, Australia

Email: yzz@maths.uq.edu.au}
\end{center}

\vskip 2cm
\begin{center}
{\bf Abstract}
\end{center}

The graded-fermion algebra and quasi-spin formalism are introduced and applied
to obtain the $gl(m|n)\downarrow osp(m|n)$ branching rules for the
``two-column" tensor irreducible representations of $gl(m|n)$, for the
case $m\leq n~ (n>2)$. In the case $m<n$, all such irreducible
representations of $gl(m|n)$ are shown to be completely reducible
as representations of $osp(m|n)$. This is also shown to be true for
the case $m=n$ except for the ``spin-singlet" representations which
contain an indecomposable representation of $osp(m|n)$ with composition
length 3. These branching rules are given in fully explicit form.

\vskip 3cm
%\noindent{\bf Mathematics Subject Classifications (1991):} 81R10, 17B37, 16W30

\end{titlepage}

%  Greek letters

\def\a{\alpha}
\def\b{\beta}
\def\d{\delta}
\def\e{\epsilon}
\def\ve{\varepsilon}
\def\g{\gamma}
\def\k{\kappa}
\def\l{\lambda}
\def\o{\omega}
\def\t{\theta}
\def\s{\sigma}
\def\D{\Delta}
\def\L{\Lambda}
\def\O{\Omega}

\def\G{{\cal G}}
\def\hG{{\hat{\cal G}}}
\def\R{{\cal R}}
\def\hR{{\hat{\cal R}}}
\def\C{{\bf C}}
\def\P{{\bf P}}
\def\Z2{{{\bf Z}_2}}
\def\T{{\cal T}}
\def\H{{\cal H}}
\def\trho{{\tilde{\rho}}}
\def\tphi{{\tilde{\phi}}}
\def\tT{{\tilde{\cal T}}}
\def\uqsnh{{U_q[\widehat{sl(n|n)}]}}
\def\uqs1h{{U_q[\widehat{sl(1|1)}]}}

% Shorthands for \begin{equation} and the like

\def\beq{\begin{equation}}
\def\eeq{\end{equation}}
\def\bea{\begin{eqnarray}}
\def\eea{\end{eqnarray}}
\def\ba{\begin{array}}
\def\ea{\end{array}}
\def\no{\nonumber}
\def\lt{\left}
\def\rt{\right}
\newcommand{\bq}{\begin{quote}}
\newcommand{\eq}{\end{quote}}

\newtheorem{Theorem}{Theorem}
\newtheorem{Definition}{Definition}
\newtheorem{Proposition}{Proposition}
\newtheorem{Lemma}{Lemma}
\newtheorem{Corollary}{Corollary}
\newcommand{\proof}[1]{{\it Proof. }
        #1\begin{flushright}$\Box$\end{flushright}}

\newcommand{\sect}[1]{\setcounter{equation}{0}\section{#1}}
\renewcommand{\theequation}{\thesection.\arabic{equation}}

\sect{Introduction}

It is well-known that 
branching rules are of great importance in the study of representation
theory. They also paly an essential role
in the determination of the parities for the components appearing in
the twisted tensor product graphs and the construction of
corresponding R-matrices \cite{Del96,Gan96}.

There appear to be virtually no results in the literature on the
branching rules for Lie superalgebras. The only exception is
ref.\cite{Van96} in which
the branching rules are determined for all typical
and atypical irreducible representations of $osp(2|2n)$ 
with respect to its subalgebra
$osp(1|2n)$. It is very interesting (and important)
to investigate the
branching rules for other Lie superalgebras. 

In this paper we investigate the anti-symmetric tensor irreducible
representations of $gl(m|n)$. This class of representations are
of interest since they are
also irreducible under the fixed point subalgebra $osp(m|n)$. 
Moreover, their quantized versions can be shown to be affinizable
to provide irreducible representations of the
twisted quantum affine superalgebra $U_q[gl(m|n)^{(2)}]$ from which
trigonometric R-matrices with $U_q[osp(m|n)]$ invariance may be
constructed  \cite{Gou99}.

These R-matrices determine new integrable models which have generated
remarkable interest in physics recently \cite{Gou97,Mar97,Sal98},
particularly in condensed matter physics where give they rise to new
integrable models of strongly correlated electrons.

To explicitly construct such R-matrices it is necessary to determine
the reduction of the tensor product of two antisymmetric tensor
irreducible representations into ``two column" irreducible
representations of $gl(m|n)$ which are then decomposed into
irreducible representations of its fixed point subalgebra $osp(m|n)$.

We determine the $gl(m|n)\downarrow osp(m|n)$
branching rules for these two column
irreducible tensor representations of
$gl(m|n)$, for the case $m\leq n,~n>2$. 
A natural framework for solving this problem is provided by 
the graded-fermion algebra 
and the quasi-spin formalism, which we introduce and develop in this paper. 
The Fock space for this graded-fermion algebra affords a convenient
realization of the class of irreducible representations of $gl(m|n)$
concerned. The reduction to $osp(m|n)$, and thus the 
$gl(m|n)\downarrow osp(m|n)$ branching rules,
can be achieved using the quasi-spin formalism.

\sect{$osp(m|n=2k)$ as a subalgebra of $gl(m|n)$}

Throughout this paper, 
we assume $n=2k$ is even and set $h=[m/2]$ so that $m=2h$
for even $m$ and $m=2h+1$ for odd $m$. 
For homogeneous operators $A, B$ we use the notation 
$[A, B]=AB-(-1)^{[A][B]}BA$ to denote the usual graded commutator.
Let $E^a_b$ be the standard generators
of $gl(m|n)$ obeying the graded commutation relations
\beq
[E^a_b, E^c_d]=\d^c_b E^a_d-(-1)^{([a]+[b])([c]+[d])}\d^a_d E^c_b.
\eeq
In order to introduce the subalgebra $osp(m|n)$ we first need a graded
symmetric metric tensor $g_{ab}=(-1)^{[a][b]}g_{ba}$ which is assumed 
to be even.
We shall make the convenient choice
\beq
g_{ab}=\xi_a\d_{a\bar{b}},
\eeq
where
\beq
\bar{a}=\left\{
\begin{array}{ll}
m+1-i,~~~& a=i\\
n+1-\mu,~~~& a=\mu,
\end{array}
\right.,~~~~~
\xi_a=\left\{
\begin{array}{ll}
1,~~~& a=i\\
(-1)^\mu,~~~& a=\mu,
\end{array}
\right..
\eeq
In the above equations, $i=1,2,\cdots,m$ and $\mu=1,2,\cdots,n$.
Note that
\beq
\xi_a^2=1,~~~\xi_a\xi_{\bar{a}}=(-1)^{[a]},~~~g^{ab}=\xi_b\d_{a\bar{b}}.
\eeq
As generators of the subalgebra $osp(m|n=2k)$ we take
\beq
\s_{ab}=g_{ac}E^c_b-(-1)^{[a][b]}g_{ac}E^c_a=-(-1)^{[a][b]}\s_{ba}
\eeq
which satisfy the graded commutation relations
\bea
[\s_{ab},\s_{cd}]&=&g_{cb}\s_{ad}-(-1)^{([a]+[b])([c]+[d])}
       g_{ad}\s_{cb}\no\\
& &-(-1)^{[c][d]}\lt(g_{bd}\s_{ac}-(-1)^{([a]+[b])([c]+[d])}
       g_{ac}\s_{db}\rt).
\eea
We have an $osp(m|n)$-module decomposition
\beq
gl(m|n)=osp(m|n)\oplus T, ~~~~[T,T]\subset osp(m|n),
\eeq
where $T$ is spanned by operators
\beq
T_{ab}=g_{ac} E^c_b+(-1)^{[a][b]}g_{bc} E^c_a=(-1)^{[a][b]}T_{ba}.
\eeq

It is convenient to introduce the Cartan-Weyl generators
\beq
\s^a_b=g^{ac}\s_{cb}=-(-1)^{[a]([a]+[b])}\xi_a\xi_b\s^{\bar{b}}_{\bar{a}}.
\eeq
As a Cartan subalgebra we take the diagonal operators
\beq
\s^a_a=E^a_a-E^{\bar{a}}_{\bar{a}}=-\s^{\bar{a}}_{\bar{a}}.
\eeq
Note that for odd $m=2h+1$ we have $\overline{h+1}=h+1$ and thus
$\s^{h+1}_{h+1}=E^{h+1}_{h+1}-E^{h+1}_{h+1}=0$.

The positive roots of $osp(m|n)$ are given by the even positive roots
(usual positive roots for $o(m)\oplus sp(n)$) together with the odd
positive roots $\d_\mu+\e_i,~1\leq i\leq m,~1\leq\mu\leq k=n/2$,
where we have adopted the useful convention
$\e_{\bar{i}}=-\e_i,~i\leq h=[m/2]$ so that $\e_{h+1}=0$ for odd
$m=2h+1$.  This is consistent with the ${\bf Z}$-gradation
\beq
osp(m|n)=L_{-2}\oplus L_{-1}\oplus L_0\oplus L_1\oplus L_2.
\eeq
Here $L_0=o(m)\oplus gl(k)$, the $gl(k)$ generators are given by
\beq
\s^\mu_\nu=E^\mu_\nu-(-1)^{\mu+\nu}E^{\bar{\nu}}_{\bar{\mu}},~~~
    1\leq \mu,\nu\leq k,
\eeq
and $L_{-2}\oplus L_0\oplus L_2=o(m)\oplus sp(n)$, where $L_2$ gives
rise to an irreducible representation of $L_0$ with highest weight
$(\dot{0}|2,\dot{0})$ spanned by the generators
\beq
\s^\mu_{\bar{\nu}}=E^\mu_{\bar{\nu}}-\xi_\mu\xi_{\bar{\nu}}E^\nu_{\bar{\mu}}=
    E^\mu_{\bar{\nu}}+(-1)^{\mu+\nu}E^\nu_{\bar{\mu}},~~~1\leq\mu,\nu\leq k.
\eeq
Finally $L_1$ is spanned by odd root space generators
\beq
\s^\mu_i=E^\mu_i+\xi_\mu E^{\bar{i}}_{\bar{\mu}}=E^\mu_i
   +(-1)^\mu E^{\bar{i}}_{\bar{\mu}},~~~1\leq\mu\leq k,~1\leq i\leq m
\eeq
and gives rise to an irreducible representation of $L_0$ with highest
weight $(1,\dot{0}|1,\dot{0})$. $L_{-1},\;L_{-2}$ give rise to
irreducible representations of $L_0$ dual to $L_1,\;L_2$, respectively.

The simple roots of $osp(m|n=2k)$ are thus given by the usual (even)
simple roots of $L_0$ together with the odd simple
root $\a_s=\d_k-\e_1$ which is the lowest weight of $L_0$-module $L_1$.
Note that the simple roots of $o(m)$ depend on whether $m$ is odd
or even, and are given here for convenience: For $m=2h$, $\a_i=\e_i-\e_{i+1}
,~~1\leq i< h,~~~\a_h=\e_{h-1}+\e_h$. For $m=2h+1$, 
$\a_i=\e_i-\e_{i+1}
,~~1\leq i< h,~~~\a_h=\e_h$.  The simple roots of $gl(k)$ are given by
\beq
\a_{h+\mu}=\d_\mu-\d_{\mu+1},~~1\leq\mu< k.
\eeq
The graded half-sum of the positive roots
of $osp(m|n=2k)$ is given by
\beq
\rho=\frac{1}{2}\sum^h_{i=1}(m-2i)\e_i+\frac{1}{2}\sum^k_{\mu=1}
    (n-m+2-2\mu)\d_\mu.
\eeq

\sect{Graded fermion realizations}

We introduce the graded anti-commutator:
\beq
\{A,B\}\equiv AB+(-1)^{[A][B]}BA.
\eeq
Note that $\{A,B\}\neq \{B,A\}$.  To realize the anti-symmetric tensor
irreducible representations of $gl(m|n)$ we introduce graded fermions $c_a$ and their
adjoints $c^\dagger_a$ obeying the graded anti-commutation relations
\beq
\{c_a,c_b\}=\{c^\dagger_a,c^\dagger_b\}=0,~~~~
   \{c_a,c^\dagger_b\}=\d_{ab}.
\eeq
Thus, when $a=i$ is even $c_i$ are fermions while for $a=\mu$ odd,
$c_\mu$ are bosons which anti-commute with the fermions.

To get a graded fermion realization of $gl(m|n)$ we set
\beq
E^a_b=c^\dagger_ac_b
\eeq
and note the graded commutation relations:
\beq
[E^a_b, c^\dagger_d]=\d_{bd}c^\dagger_a,~~~~
[E^a_b, c_d]=(-1)^{([a]+[b])[d]}\d^a_d c_b.
\eeq
Using these relations it is easy to verify that the operators
$E^a_b$ given above indeed satisfy the $gl(m|n)$ graded commutation
relations.

Thus we obtain representations of $gl(m|n)$ on the graded fermion
Fock space, which include the anti-symmetric tensor representations. The Fock
space can be shown to be completely reducible into type I unitary
irreducible representations of $gl(m|n)$ according to
\beq
F=\bigoplus^m_{a=0}\hat{V}(\dot{1}_a,\dot{0}|\dot{0})
   \bigoplus^\infty_{b=1}\hat{V}(\dot{1}|b,\dot{0}).
\eeq
Thus for $N\leq m$, the space of $N$-particle states comprises the
anti-symmetric tensor representation of $gl(m|n)$ with highest weight
$\L_N=(\dot{1}_N,\dot{0}|\dot{0})$. For $N>m$ the space of $N$
particle states comprises the irreducible representations
 of $gl(m|n)$ with highest weights
$\L_N=(\dot{1}|N-m,\dot{0})$. 

We introduce an extra ``spin" index $\a$ and consider the family of graded
fermions $c_{a\a}$ and their adjoints $c^\dagger_{a\a}$ obeying the graded
anti-commutation relations
\beq
\{c_{a\a},c_{b\b}\}=\{c^\dagger_{a\a},c^\dagger_{b\b}\}=0,~~~~
   \{c_{a\a},c^\dagger_{b\b}\}=\d_{ab}\d_{\a\b}.
\eeq
Here all spin indices are understood to be even (so that the grading
only depends on the orbital labels $a,b,c$ etc.).

We take, for our $gl(m|n)$ generators,
\beq
E^a_b=\sum_\a c^\dagger_{a\a}c_{b\a}
\eeq
which can be shown, as before, to satisfy the graded commutation relations
\beq
[E^a_b, c^\dagger_{d\a}]=\d_{bd}c^\dagger_{a\a},~~~~
[E^a_b, c_{d\a}]=(-1)^{([a]+[b])[d]}\d^a_d c_{b\a}.
\eeq
from which we deduce that the $E^a_b$ indeed obey the $gl(m|n)$ graded 
commutation relations.  Thus we may now construct more general irreducible
representations of
$gl(m|n)$ in the graded-fermion Fock space. In particular, for 
``two-column" irreducible representations, 
only two spin labels $\a=\pm$ are required.

\section{Quasi-spin (two spin labels)}

We employ the above graded-fermion algebra with two spin labels $\a=\pm$.
We set
\bea
&&Q_+=g_{dd'}c^\dagger_{d,+}c^\dagger_{d',-}=\sum_d \xi_d c^\dagger_{d,+}
      c^\dagger_{\bar{d},-},\no\\
&&Q_-=g^{dd'}c^\dagger_{d,-}c^\dagger_{d',+}=\sum_d \xi_d c_{d,-}
      c_{\bar{d},+}.
\eea
Let $Q_0=\frac{1}{2}(\hat{N}-m+n)$, where $\hat{N}=\sum^{m+n}_{a=1} E^a_a$ 
is the first order invariant of $gl(m|n)$ (i.e. the number operator). 
By straightforward computation, it can be shown that

\begin{Proposition}\label{quasi-spin alg}:
$Q_\pm, \;Q_0$ generate an $sl(2)$ Lie algebra, called
the quasi-spin Lie algebra, 
\beq
[Q_+,Q_-]=2 Q_0,~~~~[Q_0, Q_\pm]=\pm Q_\pm.
\eeq
Moreover, $Q_\pm,\;Q_0$ commute with the generators of $osp(m|n=2k)$.
\end{Proposition}

To see the significance of the graded fermion algebra for the construction
of irreducible representations we set
\beq
E^{a\a}_{b\b}=c^\dagger_{a\a}c_{b\b}
\eeq
and note the graded commutation relations
\beq
[E^{a\a}_{b\b},c^\dagger_{c\g}]=\d_{bc}\d_{\b\g}c^\dagger_{a\a},~~~~
[E^{a\a}_{b\b},c_{c\g}]=-(-1)^{[c]([a]+[b])}\d^a_c\d^\a_\g c_{b\b}
\eeq
from which we deduce 
\beq
[E^{a\a}_{b\b}, E^{c\g}_{d\d}]=\d^c_b\d^\g_\b E^{a\a}_{d\d}
  -(-1)^{([a]+[b])([c]+[d])}\d^a_d\d^\a_\d E^{c\g}_{b\b},
\eeq
which are the defining relations of $gl(2m|2n)$. That is $E^{a\a}_{b\b}$
are the generators of $gl(2m|2n)$.

As we have seen the spin averaged operators
\beq
E^a_b=\sum_{\a=\pm} E^{a\a}_{b\a}
\eeq
form the generators of $gl(m|n)$. Similarly the orbital
averaged operators
\beq
E^\a_\b=\sum_a E^{a\a}_{a\b},~~~\a,\b=\pm,
\eeq
form the generators of the spin Lie algebra $gl(2)$, which commute
with the $gl(m|n)$ generators. It is worth noting that the spin $sl(2)$
algebra with generators
\beq
S_+=E^+_-,~~~S_-=E^-_+,~~~S_0=\frac{1}{2}(E^+_+-E^-_-)\label{spin-alg}
\eeq
also commute with the quasi-spin Lie algebra. Throughout we denote the
spin Lie algebra (\ref{spin-alg})
by $sl_S(2)$ and the quasi-spin Lie algebra by $sl_Q(2)$.

Then the space of $N$ particle states gives rise to an irreducible
representation of
$gl(2m|2n)$ [and $osp(2m|2n)$] with highest weight
\beq
\lt\{
\begin{array}{ll}
(\dot{1}_N,\dot{0}|\dot{0}),~~~ & N\leq 2m\\
(\dot{1}|N-2m,\dot{0}),~~~ & N>2m.
\end{array}
\rt.
\eeq
This $N$-particle space decomposes into a multiplicity-free direct sum of
irreducible $gl(m|n)\oplus sl_S(2)$ modules
\beq
\hat{V}(a,b)\otimes V_s,
\eeq
where $V_s$ denotes the $(2s+1)$-dimensional irreducible representation
 of $sl_S(2)$,
$b=2s,~N=2a+b$ and $\hat{V}(a,b)$ denotes the irreducible representation
 of $gl(m|n)$ with highest weight
\beq
\L_{a,b}=\lt\{
\begin{array}{ll}
(\dot{2}_a,\dot{1}_b,\dot{0}|\dot{0}),~~~& a+b\leq m\\
(\dot{2}_a,\dot{1}|a+b-m,\dot{0}),~~~& a\leq m,~a+b>m\\
(\dot{2}|a+b-m,a-m,\dot{0}),~~~& a>m.
\end{array}
\rt. \label{wt-lab}
\eeq
In this way we may realize all required ``two-column" irreducible
representations of
$gl(m|n)$, inside a given anti-symmetric tensor irreducible
representation of $gl(2m|2n)$
utilising the graded-fermion calculus.

\sect{Casimir invariants and connection with quasi-spin}

{}From now on we shall use the notation
\beq
\hat{L}\equiv gl(m|n),~~~L\equiv osp(m|n),~~~\hat{L}_0\equiv gl(m)
   \oplus gl(n),~~~L_{\bar{0}}\equiv o(m)\oplus sp(n).
\eeq

Let $C_{\hat{L}},~C_L$ denote the universal Casimir invariants of
$\hat{L},~L$, respectively. Then for the
two-column irreducible representations of $\hat{L}$ we are considering, 
a straightforward but tedious 
calculation shows that
\beq
C_{\hat{L}}-C_L=(m-n+2-\frac{1}{2}\hat{N})\hat{N}-\frac{1}{2}
   (n-m)(n-m-2)+2 Q^2,\label{casimir-cc}
\eeq
where
\beq
Q^2={\bf Q}\cdot{\bf Q}=Q_0(Q_0+1)+Q_-Q_+=Q_0(Q_0-1)+Q_+Q_-\label{q^2}
\eeq
is the square of the quasi-spin.  Eq.(\ref{casimir-cc}) shows that
$Q^2$ is expressible in terms of $C_{\hat{L}},\;C_L$ and $\hat{N}$.
It follows that $Q^2,\;Q_-Q_+,\;Q_+Q_-$ must leave invariant (in fact
reduce to a scalar multiple of the identity on) a given irreducible
representation of
$L$ inside a given (two-column) representation of $\hat{L}$. Given the
highest weight of such an $L$-module we may determine its quasi-spin
$\bar{Q}$ (lowest weight of relevant $sl_Q(2)$ module) using 
(\ref{casimir-cc}) and $Q^2=\bar{Q}(\bar{Q}-1)$.

It is worth noting that we may write for our quasi-spin generators
\beq
{\bf Q}={\bf Q}^{(0)}+{\bf Q}^{(1)},
\eeq
where
\beq
Q_-^{(0)}=\sum^m_{i=1}c_{i,-}c_{\bar{i},+},~~~~ 
Q_-^{(1)}=\sum^n_{\mu=1}(-1)^\mu c_{\mu,-}c_{\bar{\mu},+}
\eeq
and similarly for $Q_+$, while
\beq
Q_0^{(0)}=\frac{1}{2}(\hat{N}_0-m),~~~~
Q_0^{(1)}=\frac{1}{2}(\hat{N}_1+n)
\eeq
with
$\hat{N}_0=\sum^m_{i=1} E^i_i$ and $\hat{N}_1=\sum^n_{\mu=1} E^\mu_\mu$
being the number operators for even fermions and odd bosons, respectively.
Then it can be shown that ${\bf Q}^{(0)},~{\bf Q}^{(1)}$
both determine $sl(2)$ algebra which commute,
so that the quasi-spin ${\bf Q}$ may be interpreted as the
total quasi-spin obtained by coupling the quasi-spins of the even and odd
components respectively.

Similar remarks apply to the total spin algebra. The total spin vector is a sum of
even and odd components
\beq
{\bf S}={\bf S}^{(0)}+{\bf S}^{(1)}
\eeq
whose corresponding $sl(2)$ algebras [c.f. (\ref{spin-alg})] are generated by
\beq
{E^{(0)}}^\a_\b=\sum^m_{i=1} E^{i\a}_{i\b},~~~~
{E^{(1)}}^\a_\b=\sum^n_{\mu=1} E^{\mu\a}_{\mu\b},
\eeq
respectively. We note that the quasi-spin and spin algebras
$sl^{(0)}_Q(2)$, $sl^{(1)}_Q(2)$, $sl^{(0)}_S(2)$, $sl^{(1)}_S(2)$ all
commute with each other.

We remark that the quasi-spin algebras $sl^{(0)}_Q(2),~sl^{(1)}_Q(2)$
play an important role in decomposing irreducible representations
of $\hat{L}_0$ into irreducible representations of $L_{\bar{0}}$. They
commute with the even subalgebra $L_{\bar{0}}$ of $L$, but not with
$L$ itself.

\sect{Quasi-spin eigenvalues}

Throughout $\hat{V}(a,b)$ denotes the irreducible representation
 of $\hat{L}$ with
highest weight $\L_{a,b}$ given by (\ref{wt-lab}).  
Let $\hat{V}_{\bar{0}}(a,b)=\hat{V}_0(\dot{0}|a+b,a,\dot{0})$ be its minimal
${\bf Z}$-graded component. 
Note that $\hat{V}_{\bar{0}}(a,b)$ is an irreducible
$gl(n)$ module and thus an irreducible $\hat{L}_0$-module. We have

\begin{Proposition}\label{cyclical}: 
$\hat{V}_{\bar{0}}(a,b)$ cyclically generates $\hat{V}(a,b)$ as an
$L$ module: viz.
\beq
\hat{V}(a,b)=U(L) \hat{V}_{\bar{0}}(a,b).
\eeq
\end{Proposition}

\noindent {\it Proof.} Set
\beq
W=U(L)\hat{V}_{\bar{0}}(a,b)\subset \hat{V}(a,b),
\eeq
i.e. $W$ is an $L$-submodule. We show equality holds. Obviously
$\hat{V}_{\bar{0}}(a,b)$ is an $L_{\bar{0}}$ module (since
$L_{\bar{0}}=L_{-2}\oplus L_0\oplus L_2\subset \hat{L}_0$). Now, since
$\hat{V}_{\bar{0}}(a,b)$ is the minimal ${\bf Z}$-graded component of
$\hat{V}(a,b)$, we have by the PBW theorem,
\beq
\hat{V}(a,b)=U(\hat{L}_+)\hat{V}_{\bar{0}}(a,b).
\eeq
Using
\beq
\s^i_\mu=E^i_\mu-(-1)^\mu E^{\bar{\mu}}_{\bar{i}}\in L_{\bar{1}}\equiv
  L_1\oplus L_{-1},
\eeq
we have
\bea
E^i_\mu \hat{V}_{\bar{0}}(a,b)&=&\s^i_\mu\hat{V}_{\bar{0}}(a,b)
    +(-1)^\mu E^{\bar{\mu}}_{\bar{i}} \hat{V}_{\bar{0}}(a,b)\no\\
&=& \s^i_\mu \hat{V}_{\bar{0}}(a,b)\subset W,
\eea
since $E^{\bar{\mu}}_{\bar{i}}\hat{V}_{\bar{0}}(a,b)\subset
\hat{L}_-\hat{V}_{\bar{0}}(a,b)=(0)$.  It follows that
\beq
\hat{L}_+\hat{V}_{\bar{0}}(a,b)\subset W.
\eeq
Proceeding recursively, let us assume that
\beq
(\hat{L}_+)^i\hat{V}_{\bar{0}}(a,b)\subset W,~~~\forall i\leq r.
\eeq
Then
\bea
E^i_\mu\hat{L}^r_+\hat{V}_{\bar{0}}(a,b)&=&\s^i_\mu\hat{L}^r_+
  \hat{V}_{\bar{0}}(a,b)+(-1)^\mu E^{\bar{\mu}}_{\bar{i}}
  \hat{L}^r_+\hat{V}_{\bar{0}}(a,b)\no\\
&\subset& L\hat{L}^r_+\hat{V}_{\bar{0}}(a,b)+\hat{L}_-
  \hat{L}^r_+\hat{V}_{\bar{0}}(a,b)\no\\
&\subset& L\hat{L}^r_+\hat{V}_{\bar{0}}(a,b)+
  \hat{L}^{r-1}_+\hat{V}_{\bar{0}}(a,b)\subset W
\eea
since $\hat{L}_-\hat{V}_{\bar{0}}(a,b)=(0)$ and  $\hat{L}^r_+
\hat{V}_{\bar{0}}(a,b)\subset W$, $\hat{L}^{r-1}_+\hat{V}_{\bar{0}}
(a,b)\subset W$ by the recursion hypothesis. Thus
$\hat{L}^{r+1}_+\hat{V}_{\bar{0}}(a,b)\subset W$
so that, by induction, $\hat{L}^r_+\hat{V}_{\bar{0}}(a,b)\subset W$,
$\forall r$. It follows that 
\beq
\hat{V}(a,b)=U(\hat{L}_+)\hat{V}_{\bar{0}}(a,b)\subset W.
\eeq
Thus we must have $W=\hat{V}(a,b)$.

{}From the traditional quasi-spin formalism for $gl(n)\supset sp(n)$ we
have a decomposition of $L_{\bar{0}}$-modules
\beq
\hat{V}_{\bar{0}}(a,b)=V_0(a,b)\oplus 
     Q^{(1)}_+\hat{V}_{\bar{0}}(a-1,b),\label{L0-decomposition}
\eeq
where $V_0(a,b)$ is an irreducible $L_{\bar{0}}$-module with highest
weight $(\dot{0}|a+b,a,\dot{0})$ and comprises quasi-spin minimal states
with respect to quasi-spin algebra ${\bf Q}^{(1)}$ (and thus also
${\bf Q}$), so
\beq
Q^{(1)}_- V_0(a,b)=Q_- V_0(a,b)=0.
\eeq

Note that for $n=2$, $\hat{V}_{\bar{0}}(a,b)=V_0(a,b)$ is
an irreducible $L_{\bar{0}}$ module, but not quasi-spin minimal. 
Thus the case $n=2$ requires a separate treatment. However 
for this case $\hat{V}_{\bar{0}}(a,b)=V_0(a,b)$ still has well-defined
quasi-spin $\bar{Q}$ (minimal weight of quasi-spin algebra): in fact
$\bar{Q}=\frac{1}{2}(b-m+n)$ for this case.

Proceeding recursively we arrive at the irreducible $sp(n)$ (and hence
$L_{\bar{0}}$) module decomposition
\beq
{\hat{V}}_{\bar{0}}(a,b)=\bigoplus^a_{c=0} {Q^{(1)}}^{a-c}_+\, V_0(c,b),
\eeq
where
\beq
{Q^{(1)}}^{a-c}_+ V_0(c,b)\cong V_0(c,b)\subset \hat{V}_{\bar{0}}(c,b)
   \label{add-1}
\eeq
is the irreducible $L_{\bar{0}}$-module with highest weight 
$(\dot{0}|c+b,c,\dot{0})$. From 
the above remarks $V_0(c,b)$ in the decomposition (\ref{add-1})
is quasi-spin minimal with 
respect to ${\bf Q}^{(1)}$ (and ${\bf Q}$) so
$Q^{a-c+1}_- {Q^{(1)}}^{a-c}_+ V_0(c,b)=(0)$.
It follows that $Q^{a+1}_-\hat{V}_{\bar{0}}(a,b)=(0)$. Thus if $q_N=
\frac{1}{2}(N-m+n)$ is the eigenvalue of $Q_0$ on $\hat{V}(a,b),~
N=2a+b$, then
\begin{Theorem}\label{allowed-Q}: 
 The quasi-spin eigenvalues (i.e. quasi-spin minimal weights) occurring in
$\hat{V}(a,b)$ lie in the range
\beq
\bar{Q}=q_N,\;q_N-1,\;\cdots,\;q_N-a,
\eeq
or $q_N\geq\bar{Q}\geq q_N-a$ (in integer steps). 
\end{Theorem}

In view of (\ref{casimir-cc}) and (\ref{q^2}) the operator 
$Q_-Q_+$ must leave invariant an $L$-submodule of $\hat{V}(a,b)$.  In
view of the above theorem the (generalized) eigenvalues of
$Q_-Q_+$ on $\hat{V}(a,b)$ must be of the form
\beq
Q_-Q_+\equiv\bar{Q}(\bar{Q}-1)-q_N(q_N+1)=(\bar{Q}+q_N)(\bar{Q}-q_N-1).
\eeq
This eigenvalue can only vanish if $\bar{Q}+q_N=0$, which would imply,
from the above theorem, $q_N-k=-q_N$ for some $0\leq k\leq a$.
Thus $k=2q_N=N-m+n$ or equivalently $a\geq N-m+n \Longleftrightarrow
a\geq 2a+b-m+n \Longleftrightarrow m-n\geq a+b$.

Thus if $m\leq n$, the (generalized) eigenvalues of $Q_-Q_+$ are all non-zero, 
except for the trivial module ($a=b=0$) which we ignore bellow. Thus
we have proved
\begin{Lemma}\label{Q-Q+}: For $m\leq n$,   $Q_-Q_+$ determines a non-singular
operator on $\hat{V}(a,b)$ except possibly for the trivial module
corresponding to $m=n,~a=b=0$.
\end{Lemma}
 
\noindent {\it Remarks}.  The above result is crucial in what follows
and will not 
generally hold for $m>n$. Hence throughout the remainder we assume
$m\leq n,~n>2$. Note that $Q_-Q_+$ is non-singular even on the trivial
module, except when $m=n$.

\sect{Induced forms and an orthogonal decomposition}

We recall that the graded fermion calculus admits a grade-$*$ operation 
defined by
\beq
(c^\dagger_{a,\a})^*=(-1)^{[a]} c_{a,\a},~~~~c^*_{a,\a}=c^\dagger_{a,\a},
\eeq
which we extend in the usual way with $(AB)^*=(-1)^{[A][B]}B^*A^*$.
This induces a grade-$*$ operation on $\hat{L}$ and $L$. Explicitly,
\beq
(E^a_b)^*=(-1)^{[a]([a]+[b])}E^b_a,~~~~
(\s^a_b)^*=(-1)^{[a]([a]+[b])}\s^b_a.
\eeq
Moreover, the quasi-spin generators satisfy
$Q^*_+=Q_-,~Q^*_-=Q_+$ and $Q^*_0=Q_0$. 

With this convention, the graded fermion Fock space admits a 
non-degenerate graded sesquilinear form $<~,~>$. In particular
$\hat{V}(a,b)$ is equipped with such a form and is non-degenerate.
Note that
\beq
<v,E^a_bw>=(-1)^{[v]([a]+[b])}<(E^a_b)^*v,w>,
\eeq
which is the invariance condition of the form. It is the unique (up
to scalar multiples) invariant graded form on $\hat{V}(a,b)$.

We now note that $Q_+\hat{V}(a-1,b)$ is an $L$-submodule of
$\hat{V}(a,b)$. In view of lemma 1 and eqs.(V.2, V.3), 
we have, 
\begin{Lemma}: The form $<~,~>$ restricted to $Q_+\hat{V}(a-1,b)\subset
\hat{V}(a,b)$ is non-degenerate except for the case $a=1,~ b=m-n=0$.
\end{Lemma}

\noindent{\it Proof.} Under the above conditions,
$Q_-Q_+$ is non-singular on $\hat{V}(a-1,b)$, so
$Q_-Q_+\hat{V}(a-1,b)=\hat{V}(a-1,b)$. 
Hence for $v\in\hat{V}(a-1,b)$, we have
$0=<Q_+\hat{V}(a-1,b),Q_+v>$ $~\Longrightarrow~$
$0=<Q_-Q_+\hat{V}(a-1,b),v>=<\hat{V}(a-1,b),v>~\Longrightarrow ~ v=0$ 
since $<~,~>$ on 
$\hat{V}(a-1,b)$ is non-degenerate. This shows that the form $<~,~>$
restricted to $Q_+\hat{V}(a-1,b)$ is non-degenerate as required.

In view of proposition \ref{cyclical}, we have
\begin{Proposition}\label{Q-V=V}:   $Q_-\hat{V}(a,b)=\hat{V}(a-1,b)$.
\end{Proposition}

\noindent{\it Proof.} From proposition \ref{cyclical} we have
\bea
Q_-\hat{V}(a,b)&=&Q_- U(L)\hat{V}_{\bar{0}} (a,b)=U(L)Q_-\hat{V}
  _{\bar{0}} (a,b)\no\\
&=&U(L) Q^{(1)}_- \hat{V}_{\bar{0}} (a,b)=U(L)\hat{V}_{\bar{0}} (a-1,b),
\eea
where the last step follows from a clasical Lie algebra result.
Again utilising proposition \ref{cyclical}
 we have $U(L)\hat{V}_{\bar{0}} (a-1,b)
=\hat{V}(a-1,b)$ from which the result follows.

We are now in a position to prove
\begin{Proposition}\label{L-decomposition}: 
We have an $L$-module orthogonal decomposition
\beq
\hat{V}(a,b)={\cal K}\oplus Q_+\hat{V}(a-1,b),
\eeq
where ${\cal K}={\rm Ker}Q_-\bigcap\hat{V}(a,b)$, except for the case
$a=1,~ b=m-n=0$.
\end{Proposition}

\noindent{\it Proof.} For $v\in\hat{V}(a,b)$, $<v,Q_+\hat{V}(a-1,b)>=0
\Longleftrightarrow <Q_-v,\hat{V}(a-1,b)>=0 \Longleftrightarrow
Q_-v=0$ (by proposition \ref{Q-V=V}) $\Longleftrightarrow  v\in{\cal K}$.
Since $<~,~>$ restricted to $Q_+\hat{V}(a-1,b)$ is non-degenerate,
the result follows.

Finally, in view of theorem \ref{allowed-Q} we have
\begin{Proposition}: $\hat{V}(a=0,b)$ is an irreducible $L$-module.
\end{Proposition}

\noindent{\it Proof.} In such a case $\hat{V}_{\bar{0}}(0,b)=V_0(0,b)$ is
an irreducible $L_{\bar{0}}$-module cyclically generated by an $L$
maximal state. Thus $\hat{V}(0,b)=U(L)V_0(0,b)$ must be an
indecomposable $L$-module. Since the form $<~,~>$ on $\hat{V}(0,b)$ is
non-degenerate, this forces $\hat{V}(0,b)$ to be an irreducible 
$L$-module.

The result above shows that the minimal $\hat{L}$ irreducible
representations are indeed 
irreducible under $L$.

\sect{Preliminaries to branching rules}

It is our aim below to prove, barring the exceptional case of lemma 2,
that ${\cal K}$ is an irreducible $L$ module.
Note that the maximal state of the $L_{\bar{0}}$ module $V_0(a,b)$
occurring in the decomposition (\ref{L0-decomposition}) in fact coincides
with the $\hat{L}_0$ maximal vector $v^\L_+$ of $\hat{V}_{\bar{0}}(a,b)$:
For $n>2$ it can be seen directly that
\beq
Q_-v^\L_+=Q^{(1)}_- v^\L_+=0
\eeq
for this maximal vector. Moreover for $n>2$ we have
\beq
E^i_{\bar{\mu}} v^\L_+=0,~~~~  1\leq i\leq m,~ 1\leq\mu\leq k
\eeq
otherwise this vector would have weight 
$(\dot{0}|a+b,a,\dot{0})+\e_i-\d_{\bar{\mu}}$~ ($\bar{\mu}>k=n/2$),
which is impossible since all $\hat{L}$ weight components are positive.
Also, since $v^\L_+$ belongs to the $\hat{L}$ minimal ${\bf Z}$-graded
component, we must have
\beq
E^\mu_i v^\L_+=0,~~~\forall i,\mu.
\eeq
Thus for $\s^\mu_i\in L_1$ we have
\bea
&&\s^\mu_i v^\L_+=(E^\mu_i+(-1)^\mu E^{\bar{i}}_{\bar{\mu}})
   v^\L_+=0,~~~\forall i,~1\leq\mu\leq k\no\\
&&~~~\Longrightarrow ~~L_1 v^\L_+=(0).
\eea

It follows that the $L_{\bar{0}}$ module $V_0(a,b)$ must cyclically
generate an indecomposable module over $L$:
\beq
V(a,b)=U(L) V_0(a,b)\label{cyclical-L}
\eeq
with highest weight 
\beq
\l_{a,b}\equiv (\dot{0}|a+b,a,\dot{0}).
\eeq
 Since
\beq
Q_- V_0(a,b)=Q_-^{(1)} V_0(a,b)=(0)
\eeq
we have
\beq
Q_- V(a,b)=Q_- U(L) V_0(a,b)=U(L)Q_- V_0(a,b)=(0).
\eeq
It follows that $V(a,b)\subset {\cal K}$.
 
We now show that $V(a,b)={\cal K}$ is irreducible. First, in view of
proposition \ref{Q-V=V}, we have
\begin{Lemma}\label{v in K}:  $v\in {\cal K}\Longleftrightarrow Q_+Q_-v=0$.
\end{Lemma}

\noindent{\it Proof.} Obviously $v\in{\cal K} \Longrightarrow Q_-v=0
\Longrightarrow Q_+Q_-v=0$. Conversely, $Q_+Q_-v=0 \Longrightarrow$
\beq
0=<Q_+Q_-v,\hat{V}(a,b)>=<Q_-v,Q_-\hat{V}(a,b)>=<Q_-v, \hat{V}(a-1,b)>\no
\eeq
$\Longrightarrow Q_-v=0 \Longrightarrow v\in{\cal K}$.

It follows that ${\cal K}$ consists of eigenstates of $Q_+Q_-$ with zero
eigenvalue. Also since $Q_-{\cal K}=(0)$ and ${\cal K}\subset\hat{V}(a,b)$
it follows that all states in ${\cal K}$ are eigenvectors of $Q_0$ with
eigenvalue $q_N=\frac{1}{2}(N-m+n)$ and are moreover quasi-spin
minimal states and so have quasi-spin $\bar{Q}=q_N$. Thus $Q^2$ reduces
to a scalar multiple $\bar{Q}(\bar{Q}-1)=q_N(q_N-1)$ on ${\cal K}$.
It then follows from (\ref{casimir-cc}) that the universal Casimir 
element $C_L$ of $L$ must reduce to a scalar multiple of the identity
on ${\cal K}$. Since $V(a,b)\subset{\cal K}$ has highest weight
$\l_{a,b}$, this eigenvalue can be shown to be given by
\beq
\chi_{\l_{a,b}}(C_L)=(\l_{a,b},\l_{a,b}+2\rho)
   =-(a+b)(a+b+n-m)-a(a+n-m-2).\label{chi-C}
\eeq
Hence we have proved
\begin{Lemma}\label{chi-CL}: $C_L$ reduces to a scalar multiple of the
identity on ${\cal K}$ with eigenvalue given by (\ref{chi-C}).
\end{Lemma}

Now ${\cal K}$ is a completely reducible $L_{\bar{0}}$-module. Hence

\begin{Lemma}\label{irre K}: Suppose for any irreducible $L_{\bar{0}}$
module $V_0(\l)$ contained in an irreducible $\hat{L}_0$ module
 $\hat{V}_0({\L})\subset\hat{V}(a,b)$ that:
$\chi_\l(C_L)=\chi_{\l_{a,b}}(C_L) \Longleftrightarrow
{\L}=\L_{a,b}$ and $\l=\l_{a,b}$. Then ${\cal K}=V(a,b)$ is
irreducible.
\end{Lemma}

\noindent{\it Proof.} Indeed in such a case it follows from lemma
\ref{chi-CL}, that the highest weight vector of $V(a,b)$ must be the 
unique primitive vector in ${\cal K}$. This is enough to prove that
${\cal K}$ is irreducible.

Finally, we recall that
$\hat{V}(a,b)$ comprises states with total spin $s=b/2$ and with particle
number $N=2a+b$. Then the possible irreducible representations
 of $\hat{L}_0$
occurring in $\hat{V}(a,b)$ must have highest weights of the form
\beq
{\L}=(\dot{2}_{a'},\dot{1}_{b'},\dot{0}|c',d',\dot{0}).\label{add-2}
\eeq
Then we must have
\beq
2a'+b'+c'+d'=N=2a+b.
\eeq
Moreover the total spins for the even and odd components of this irreducible
representation are
$s_0=b'/2$ and $s_1=(c'-d')/2$, respectively. So, using the triangular rule for
angular momenta, we have
\beq
s\leq s_0+s_1,~~~~s_0\leq s+s_1,~~~~s_1\leq s+s_0
\eeq
or
\beq
b\leq b'+c'-d',~~~b'\leq b+c'-d',~~~c'-d'\leq b+b'.\label{inequality}
\eeq
These inequalities turn out to be important bellow.

\sect{$\hat{L}\downarrow L$ branching rules}

We start this section with some facts concerning $\hat{L}_0
\downarrow L_{\bar{0}}$. The possible $\hat{L}_0$ highest weights $\L$
occuring in $\hat{V}(a,b)$ are of the form of (\ref{add-2}).
The possible $L_{\bar{0}}$ highest weights $\l$ in
$\hat{V}(a,b)$ are obtained from such
${\L}$ by a classical contraction procedure and have the form
\beq
\l=(\dot{2}_c,\dot{1}_d,\dot{0}|e,f,\dot{0}),~~~~c+d\leq h,
\eeq
where $d=b'\wedge (m-2c-b'),~e-f=c'-d'$ [here and below
$x\wedge y\equiv {\rm min}(x,y)$] and 
\beq
c\leq a',~ ~~~e+f\leq c'+d' = 2a+b-2a'-2b'.
\eeq
Note that for $n>4$, there are additional restrictions on the allowed
$L_{\bar{0}}$ dominant weights in order that they give rise to highest
weights of $L$ \cite{Kac78}. In the interests of a unified treatment
of all cases, including $n=4$, we do not impose these supplementary
conditions here.

Since $e-f=c'-d'$ the
inequalities (\ref{inequality}) lead to
\beq
b'\leq b+e-f,~~~b\leq b'+e-f,~~~e-f\leq b+b'.
\eeq
Hence we have the inequalities
\begin{Lemma}\label{e< f<}: $e\leq a+b-c,~~ f\leq a-c$.
\end{Lemma}

\noindent{\it Proof.} We have
\bea
e+f\leq 2a+b-2a'-b',~~~~e-f\leq b+b'.\no
\eea
Adding these two inequalities
 gives $e\leq a+b-a'$. Thus $e\leq a-c$ since $c\leq a'$.
Similarly, adding
\bea
e+f\leq 2a+b-2a'-b',~~~~f-e\leq b'-b\no
\eea
leads to $f\leq a-a'\leq a-c$.

We are now in a position to compute the eigenvalue 
$\chi_\l(C_L)$ compared
with that of (\ref{chi-C}). By direct computation we have
\bea
\chi_\l(C_L)&=&(\l,\l+2\rho)=m(2c+d)-c(c+1)-(c+d)(c+d+1)\no\\
& & -(n-m)(e+f)+4c+d
   +2f-e^2-f^2,\label{chi-C2}
\eea
where we have used
\beq
\l=\sum^c_{i=1}2\e_i+\sum^{d+c}_{i=c+1}\e_i+e\d_1+f\d_2
\eeq
together with the expression for $\rho$ of $L$. By a straightforward but
tedious calculation, using (\ref{chi-C}) and (\ref{chi-C2}), we obtain
\bea
\chi_\l(C_L)-\chi_{\l_{a,b}}(C_L)&=&2cn+d(m-d)+2c(2a+b-2c-d)\no\\
& &+(a+b-c-e)(a+b-c+e+n-m)\no\\
& &+(a-c-f)(a-c+f+n-m-2)\label{chi-chi1}\\
&=&[2c(n+1)+2f-2a]+d(m-d)+2c(2a+b-2c-d)\no\\
& &+(a+b-c-e)(a+b-c+e+n-m)\no\\
& &+(a-c-f)(a-c+f+n-m).\label{chi-chi2}
\eea
All terms on the r.h.s. of (\ref{chi-chi1}) are positive, in view of the
inequalities given above, except possibly the last due to the term
$(a-c+f+n-m-2)$. Similarly in (\ref{chi-chi2}) all terms on the r.h.s.
are positive except possibly the first.

We proceed step wise.
\vskip.1in
\noindent{\bf (i)} $\underline{c\geq 1}$: ~~ Then the first term on the r.h.s.
of (\ref{chi-chi2}) gives
\bea
2c(n+1)+2f-2a\geq 2(n+1+f-a).\no
\eea
This leads to two subclasses: 
\vskip.1in
\noindent{\bf (i.1)}
 $\underline{a\leq n+1}$: ~ The the r.h.s. terms are all non-negative,
so (\ref{chi-chi2}) can only vanish if $a=n+1,~f=0=d,~2a+b=2c+d$.
But then, since $d=0$ this would imply $2c=2a+b \Longrightarrow c\geq 
a=n+1$, which is impossible since $c\leq h\leq m\leq n$. Thus we conclude that the 
r.h.s. must be strictly positive in this case.
\vskip.1in
\noindent{\bf (i.2)}
 $\underline{a\geq n+2}$:~ In this case all terms on the r.h.s. of
(\ref{chi-chi1}) are non-negative including the last term since, for the
case at hand,
\bea
a-c+f+n-m-2&\geq& n+2-c+f+n-m-2\no\\
&\geq& n-c+f+n-m\geq 0\no
\eea
since $n\geq m\geq h\geq c$. Since $c\geq 1$, the r.h.s. of (\ref{chi-chi1})
must be strictly positive in this case.

We thus conclude, for $c\geq 1$, that $\chi_\l(C_L)-\chi_{\l_{a,b}}(C_L)
>0$. It remains then to consider the case $c=0$ in which case we have
\bea
\chi_\l(C_L)-\chi_{\l_{a,b}}(C_L)& =&d(m-d)+(a+b-e)(a+b+e+n-m)\no\\
 & &   +(a-f)(a+f+n-m-2).\label{chi-chi3}
\eea
Note that for the case $c=0$, the inequalities of lemma \ref{e< f<}
reduce to $e\leq a+b,~ f\leq a$ and for the case at hand we have
\bea
e-f=c'-d',~~~~d=b'\wedge (m-b').\no
\eea
It is convenient to treat the cases $m=n$ and $m<n$ separately.
\vskip.1in
\noindent{\bf (ii)}
 $\underline{c=0,~n>m}$:~~ Here we assume $a\geq 1$, since when
$a=0$, $\hat{V}(a=0,b)$ is already known to be an irreducible $L$ module,
so the branching rule is trivial.

Under these assumptions all terms on the r.h.s. of (\ref{chi-chi3}) are
non-negative, including the last since
\bea
a+f+n-m-2\geq f+n-m-1\geq 0.\no
\eea
Note that this factor can only vanish when $a=1,~f=0,~n=m+1$. There
are thus two possibilities to consider for vanishing of the r.h.s. of
(\ref{chi-chi3}):
\vskip.1in
\noindent{\bf (ii.1)} $\underline{d=0,~e=a+b,~f=a}$: ~ Since $c'+d'=2a+b-2a'-b'
\geq e+f=2a+b$ and $c'-d'=e-f=b$,
this implies $a'=b'=0,~c'=a+b,~d'=a$ and $\l=\l_{a,b}$. So in this case
${\L}=(\dot{0}|a+b,a,\dot{0})=\L_{a,b}$ and $\l=\l_{a,b}$.
\vskip.1in
\noindent{\bf (ii.2)} $\underline{d=0,~e=a+b,~f=0,~a=1,~n=m+1}$: ~ 
Then $c'+d'\geq e+f=a+b$. Since $a=1$ we thus have
\bea
2+b=N=2a'+b'+c'+d'\geq 2a'+b'+a+b=2a'+b' +1+b.\no
\eea
$\Longrightarrow 1\geq 2a'+b' \Longrightarrow a'=0$ and $b'\leq 1$. In such
a case we must have $d=b'\wedge (m-b')$ and since $d=0
\Longrightarrow b'=0$, or $m=b'=1~\Longrightarrow n=2$ which we ignore.
Then ${\L}=(\dot{0}|c',b',\dot{0})$ with
$c'-b'=e-f=a+b=1+b$ which corresponds to states with spin $(1+b)/2$
which is impossible since all states in $\hat{V}(a,b)$ have spin $b/2$.
Thus this latter case can not occur.

Thus we have shown, for all cases, that when $n>m$, ${\cal K}=V(a,b)$
must be an irreducible module with highest weight $\l_{a,b}$, using
lemma \ref{chi-CL}.

In view of proposition \ref{Q-V=V} we thus have the $L$ module decomposition
\beq
\hat{V}(a,b)=V(a,b)\oplus Q_+\hat{V}(a-1,b).\label{vhat-decom}
\eeq
Since $Q_-Q_+$ is non-singular, $Q_+\hat{V}(a-1,b)\cong\hat{V}(a-1,b)$.
By repeated application of (\ref{vhat-decom}) we arrive at the
irreducible $L$ module decomposition
\beq
\hat{V}(a,b)=\bigoplus^a_{c=0}Q^{a-c}_+ V(c,b).
\eeq
Hence we have proved 
\begin{Theorem}\label{branching rule1} ($n>m,~n>2$): 
We have the irreducible $L$-module
decomposition
\beq
\hat{V}(a,b)=\bigoplus^a_{c=0} V(c,b).
\eeq
\end{Theorem}

We emphasize that throughout $V(a,b)$ denotes the $L$-module with
highest weight $\l_{a,b}=(\dot{0}|a+b,a,\dot{0})$.
It remains now to consider the case $m=n$ which is somewhat more interesting.
\vskip.1in
\noindent{\bf (iii)}
 $\underline{c=0,~m=n>2}$:~~  Again we assume $a\geq 1$ since
$\hat{V}(a=0,b)$ is an irreducible $L$ module as we
have seen. We recall for the case at hand
$e\leq a+b$, $f\leq a$, $a\geq 1$, $m=n>2$, $e-f=c'-d'$, $d=b'\wedge
(m-b')$ and
\bea
\chi_\l(C_L)-\chi_{\l_{a,b}}(C_L)&=&d(m-d)+(a+b-e)(a+b+e)\no\\
& & +(a-f) (a+f-2).\label{chi-chi4}
\eea
There are now several cases to consider for vanishing of (\ref{chi-chi4}).
\vskip.1in
\noindent{\bf (iii.1)} $\underline{a=f}$: ~ Then (\ref{chi-chi4}) vanishes when
$d=0,~ e=a+b$. Thus
\bea
c'+d'\geq e+f=2a+b=2a'+b'+c'+d'\no
\eea
$\Longrightarrow a'=b'=0, ~ c'+d'=2a+b$ and $c'-d'=e-f=b$. This
corresponds to ${\L}=\L_{a,b}$ and $\l=\l_{a,b}$.
\vskip.1in
\noindent{\bf (iii.2)}
 $\underline{f=2-a}$: ~ Then (\ref{chi-chi4}) vanishes when
$d=0,~ e=a+b$. Since $a\geq 1$ there are two cases:\\
{\bf (iii.2.1)} ${f=0,~a=2}$:~ This is only possible when
$c'+d'\geq e+f=a+b$ $\Longrightarrow$
\bea
2a+b\geq 2a'+b'+c'+d'\geq 2a'+b'+a+b\no
\eea
$\Longrightarrow a\geq 2a'+b'$ or $2\geq 2a'+b'$.  This leads to two further
cases:\\
{\bf (iii.2.1a)} ${f=0,~a=2,~ a'=0,~ b'\leq 2}$: ~ In view of the
contraction procedure this is only consistent with $d=0$ if $b'=0$
(so $ c'=a+b,~ d'=a$) or if $b=2$ and $m=n=2$. The latter case is being
ignored and the former case can not occur since then $c'-d'=e-f=a+b>b$ in
contradiction to the fact that all states in $\hat{V}(a,b)$ have spin
$b/2$.\\
{\bf (iii.2.1b)} ${f=0,~a=2,~a'=1,~b'=d=0}$:~ Then $c'-d'=e-f=a+b>b$
which again is impossible since all states have spin $b/2$.
%\vskip.1in
\\
{\bf (iii.2.2)} ${f=a=1}$: ~Then $c'-d'=a+b-a=b$, $c'+b'
\geq e+f=2a+b$ $\Longrightarrow a'=b'=0,~ c'=a+b,~ d'=a$
$\Longrightarrow$ ${\L}=\L_{a,b},~ \l=\l_{a,b}$.
\vskip.1in
\noindent{\bf (iii.3)} $\underline{a+f-2<0,~ a>f}$:~ This can only occur when 
$a=1,~f=0$ in which case the r.h.s. of (\ref{chi-chi4}) becomes
\bea
d(m-d)+(a+b+e)(a+b-e)-1.\no
\eea
There are two cases for the vanishing of this:\\
{\bf (iii.3.1)}
 $e=a+b,~d=1,~m=2$ which can occur but we are ignoring since $n=m>2$.
\\
{\bf (iii.3.2)} $ d=f=e=b=0$:~ Then $c'-d'=e-f=0$ and
\bea
N=2=2a+b=2a'+b'+c'+d'=2(a'+c')+b'\no
\eea
which can occur in the following cases:
\bea
&&a'=b'=0,~~~c'=d'=1~\Longrightarrow ~\l=(\dot{0}|\dot{0}),~~
   {\L}=(\dot{0}|1,1,\dot{0});\no\\
&&b'=c'=d'=0,~~~a'=1~
  \Longrightarrow~ \l=(\dot{0}|\dot{0}),~~
   {\L}=(2,\dot{0}|\dot{0}).\no
\eea
\vskip.1in
This exhausts all possibilities. It follows from the above that for
$n=m>2$ the r.h.s. of (\ref{chi-chi4}) is always strictly positive
and can only vanish in the last case, corresponding to
$a=1$ and $b=0$. This is the irreducible representation
 $\hat{V}(2,\dot{0}|\dot{0})$ of $gl(n|n)$
which is known to give rise to an indecomposable $osp(n|n)$
module with a composition series of length 3 whose factors are
isomorphic to the $osp(n|n)$ modules $V(1,0)$ and $V(0,0)$ (see Appendix
A).

Thus we have proved the decomposition
\beq
\hat{V}(a,b)=V(a,b)\bigoplus Q_+\hat{V}(a-1,b)\label{vhat-decom2}
\eeq
with $V(a,b)$ an irreducible $L$-module of highest weight $\l_{a,b}$,
provided $(a,b)\neq (1,0)$. Proceeding recursively we have
\begin{Theorem}\label{branching rule2} ($n=m>2$): For $b>0$ we have the
irreducible $L$-module decomposition
\beq
\hat{V}(a,b)=\bigoplus^a_{c=0}V(c,b).
\eeq
For $b=0$ we have the $L$-module decomposition
\beq
\hat{V}(a,0)=\bigoplus^a_{c=1}V(c,0),
\eeq
where $V(c,0)$ is irreducible for $c>1$ but $V(1,0)$ is indecomposable
with a composition series of length 3 with composition factors
isomorphic to irreducible $L$-modules $V(1,0)$ and $V(0,0)$, the latter
occuring twice.
\end{Theorem}

Theorems \ref{branching rule1} and \ref{branching rule2} are our main
results in this section concerning the $\hat{L}\downarrow L$ branching rules
for the two-column tensor represetations of $\hat{L}$. We remark that
for the special case $n-m=0=b,~a=1$,  $\hat{V}(a-1,b)=\hat{V}(0,0)$
coincides with the identity module which is the exceptional case of 
lemma 2. For this case the
form $<~,~>$ on $\hat{V}(a,b)=\hat{V}(1,0)$ is degenerate on
$Q_+\hat{V}(a-1,b)=Q_+\hat{V}(0,0)$. Thus proposition 
\ref{L-decomposition}
fails in this case (and only this case). This of course agrees with
the result that $\hat{V}(a,b)=\hat{V}(1,0)\equiv\hat{V}(2,\dot{0}|\dot{0})$
is indecomposable for $m=n$.

\vskip.3in
\noindent {\it Acknowledgements.} 
This paper was completed when YZZ visited Northwest University, China.
He thanks Australian Research Council IREX programme for an Asia-Pacific
Link Award and Institute of Modern Physics of the Northwest University
for hospitality. The financial support from Australian Research 
Council large, small and QEII fellowship grants is also gratefully acknowledged.

\appendix

\sect{Appendix}

Here for completeness we determine the structure of the irreducible
$\hat{L}=gl(n|n=2k)$ module $\hat{V}(2,\dot{0}|\dot{0})$ as a module
over $L=osp(n|n)$, in fully explicit form.

First $\hat{V}(2,\dot{0}|\dot{0})$ admits the following ${\bf Z}$-graded
decomposition into irreducible $\hat{L}_0$-modules with highest weights
shown:
\bea
\hat{V}(2,\dot{0}|\dot{0})=\hat{V}_0(2,\dot{0}|\hat{0})\oplus
  \hat{V}_1(1,\dot{0}|1,\dot{0})\oplus\hat{V}_2(\dot{0}|1,1,\dot{0}).\no
\eea
In the notation of the paper, the last space corresponds to the
irreducible $\hat{L}_0$-module $\hat{V}_{\bar{0}}(a=1,b=0)$. In terms
of the graded fermion formalism we have the following basis states:
\bea
\hat{V}_0(2,\dot{0}|\dot{0})&:& ~~(c^\dagger_{i,+}c^\dagger_{j,-}
    +c^\dagger_{j,+}c^\dagger_{i,-})|0>,~~~1\leq i,j\leq n,\no\\
\hat{V}_1(1,\dot{0}|1,\dot{0})&:& ~~(c^\dagger_{i,+}c^\dagger_{\mu,-}
    +c^\dagger_{\mu,+}c^\dagger_{i,-})|0>,~~~1\leq i,\mu\leq n,\no\\
\hat{V}_2(\dot{0}|1,1,\dot{0})&:& ~~(c^\dagger_{\mu,+}c^\dagger_{\nu,-}
    -c^\dagger_{\nu,+}c^\dagger_{\mu,-})|0>,~~~1\leq \mu,\nu\leq n,
     \label{appendix-1}
\eea
where $|0>$ is the vacuum state. The latter space decomposes into
$L_{\bar{0}}$-modules according to
\bea
\hat{V}_2(\dot{0}|1,1,\dot{0})=V_0(\dot{0}|1,1,\dot{0})\oplus V_0
   (\dot{0}|\dot{0}),\no
\eea
where $V_0(\dot{0}|\dot{0})$ is spanned by $Q^{(1)}_+|0>$ 
(the trivial $L_{\bar{0}}$-module) and $V_0(\dot{0}|1,1,\dot{0})$ is
an irreducible $L_{\bar{0}}$-module with highest weight indicated and
the following basis vectors:
\bea
&& (c^\dagger_{\mu,+}c^\dagger_{\nu,-}
    -c^\dagger_{\nu,+}c^\dagger_{\mu,-})|0>,~~~1\leq \nu\neq\bar{\mu}\leq n,
   \label{appendix-2}\\
&&
(\O^\dagger_\mu-\O^\dagger_{\mu+1})|0>,~~~1\leq\mu<k,\label{appendix-3}
\eea
where
\bea
\O_\mu^\dagger\equiv c^\dagger_{\mu,+}c^\dagger_{\bar{\mu},-}
   -c^\dagger_{\bar{\mu},+}c^\dagger_{\mu,-}.\no
\eea
Note that this irreducible $L_{\bar{0}}$ module cyclically generates
an indecomposable $L$-module $\tilde{V}(\d_1+\d_2)$ with highest
weight $\d_1+\d_2$ and highest weight vector given by (\ref{appendix-2})
with $\mu=1,~\nu=2$.

Now $\hat{V}_1(1,\dot{0}|1,\dot{0})$ is also irreducible as an
$L_{\bar{0}}$-module which is contained in $\tilde{V}(\d_1+\d_2)$.
Then by applying the odd lowering generators 
$\s^i_\mu=E^i_\mu-(-1)^\mu E^{\bar{\mu}}_{\bar{i}}~(1\leq \mu \leq k,~
1\leq i\leq n)$ of $L$ to the states (\ref{appendix-1}), the following
states in $\hat{V}_0(2,\dot{0}|\dot{0})$ are easily seen to be in
$\tilde{V}(\d_1+\d_2)$:
\bea
&& (c^\dagger_{i,+}c^\dagger_{j,-}
    +c^\dagger_{j,+}c^\dagger_{i,-})|0>,~~~1\leq j\neq\bar{i}\leq n,
   \label{appendix-4}\\
&&
(\O^\dagger_i-\O^\dagger_{i+1})|0>,~~~1\leq i<k,\label{appendix-5}
\eea
where
\bea
\O_i^\dagger\equiv c^\dagger_{i,+}c^\dagger_{\bar{i},-}
   +c^\dagger_{\bar{i},+}c^\dagger_{i,-}.\no
\eea
Further the following states are aslo seen to be in
$\tilde{V}(\d_1+\d_2)$:
\bea
(\O^\dagger_i+(-1)^\mu\O^\dagger_\mu)|0>,~~~1\leq i, \mu<k\label{*}
\eea
which follows by applying $\s^{\bar{\mu}}_i$ to the states
(\ref{appendix-1}) with $1\leq\mu\leq k$. Summing (\ref{*}) on $\mu=i$
from 1 to $k$ we thus obtain
\bea
\lt(\sum_{i=1}^k\O^\dagger_i+\sum_{\mu=1}^k(-1)^\mu\O^\dagger_\mu\rt)|0>
   =Q_+|0>\in \tilde{V}(\d_1+\d_2).\label{appendix-6}
\eea
It is worth noting that the states (\ref{*}) are expressible in terms of
the states (\ref{appendix-3}), (\ref{appendix-5}) and
(\ref{appendix-6}).

The states (\ref{appendix-1} -- \ref{appendix-6}) form a basis for the
standard cyclic $L$-module $\tilde{V}(\d_1+\d_2)$. We note that
${\rm dim}\tilde{V}(\d_1+\d_2)={\rm dim}\hat{V}(2,\dot{0}|\dot{0})-1$
and $\tilde{V}(\d_1+\d_2)$ is the unique maximal $L$-submodule of
$\hat{V}(2,\dot{0}|\dot{0})$. In view of (\ref{appendix-6}), this
module is not irreducible since it contains the trivial one-dimensional
$L$-module $V(\dot{0}|\dot{0})$ as a unique submodule.

The remaining state in $\hat{V}(2,\dot{0}|\dot{0})$, not in
$\tilde{V}(\d_1+\d_2)$, is $Q^{(1)}_+|0>$ (or $Q^{(0)}_+|0>$) which
thus generates the basis vector for the $L$ factor module
$\hat{V}(2,\dot{0}|\dot{0})/ \tilde{V}(\d_1+\d_2)$ which is obviously
isomorphic to the trivial $L$-module $V(\dot{0}|\dot{0})$. We thus
arrive at the $L$-module composition series
$\hat{V}(2,\dot{0}|\dot{0})\supset \tilde{V}(\d_1+\d_2)\supset
   V(\dot{0}|\dot{0})\supset (0)$
with corresponding factors isopmorphic to the irreducible $L$-modules
with highest weights $(\dot{0}|\dot{0}),~ \d_1+\d_2$ and
$(\dot{0}|\dot{0})$, respectively.

This result is of importance to the explicit construction of new
R-matrices \cite{Gou99}. In particular it gives rise to an
$L$-invariant nilpotent contribution to the R-matrices, a new effect
not seen in the untwisted or non-super cases.

%\newpage


\vskip.3in
\begin{thebibliography}{99}
\bibitem{Del96} G.W. Delius, M.D. Gould, Y.-Z. Zhang, Int. J. Mod. Phys.
   {\bf A11}, 3415 (1996).
\bibitem{Gan96} G.M. Gandenberger, N.J. MacKay, G.M.T. Watts,
   Nucl. Phys. {\bf B465}, 329 (1996).
\bibitem{Van96} J. Van der Jeugt, J. Math. Phys. {\bf 37}, 4176 (1996).
\bibitem{Gou99} M.D. Gould, Y.-Z. Zhang, {\it Twisted quantum affine
   superalgebra $U_q[gl(m|n)^{(2)}]$ and new $U_q[osp(m|n)]$ invariant
   R-matrices}, preprint to appear in math-QA.
\bibitem{Gou97} M.D. Gould, J.R. Links, I. Tsohantjis, Y.-Z. Zhang,
   J. Phys. {\bf A30}, 4313 (1997).
\bibitem{Mar97} M.J. Martins, P.B. Ramos, Phys. Rev. {\bf B56}, 6376
   (1997).
\bibitem{Sal98} H. Saleur,  {\it The long delayed solution of the Bukhvostov-Lipatov
    model}, e-print hep-th/9811023.
\bibitem{Kac78} V.G. Kac, Lect. Notes in Math. {\bf 676}, 597 (1978).
\end{thebibliography}
\end{document}